\documentclass[twocolumn,english,aps,prl,showpacs]{revtex4-1}
\usepackage{amssymb}
\usepackage{amsmath}
\usepackage[pdftex]{graphicx}

\begin{document}

\title{Quasiperiodicity and 2D Topology in 1D Charge Ordered Materials}

\author{Felix Flicker}
  \email{flicker@physics.org}
  \affiliation{H. H. Wills Physics Laboratory, University of Bristol, Tyndall Avenue, Bristol, BS8 1TL, UK}
\author{Jasper van Wezel} 
  \affiliation{H. H. Wills Physics Laboratory, University of Bristol, Tyndall Avenue, Bristol, BS8 1TL, UK}

\begin{abstract}
It has recently been argued that individual 1D quasicrystals can be ascribed 2D topological quantum numbers and a corresponding set of topologically protected 
edge modes. Here, we demonstrate the equivalence of such 1D quasicrystals to a mean-field treatment of incommensurate charge order in 1D materials. 
Using the fractal nature of the spectrum of commensurate charge-ordered states we consider incommensurate order as a limiting case of commensurate orders. We 
show that their topological properties arise from a 2D parameter space spanned by both phase and wave vector, bringing the observation of 2D edge modes in line 
with the standard classification of topological order. 
 The equivalence also provides a set of real-world quasiperiodic materials which can be readily experimentally examined. 
We propose an experimental test of both the quasicrystalline and topological character of these systems in the form of a quantized adiabatic particle transport 
upon dragging the charge-ordered state.
\end{abstract}

\maketitle

The second half of the twentieth century saw two revolutions in condensed matter physics. The first began with the mathematical observation that solids need 
not involve periodic repetitions of unit cells, but can instead involve aperiodic tilings of two or more inequivalent cells~\cite{SnS2}. 
Experimental verification in the form of a `quasicrystalline' Al-Mn alloy followed shortly afterwards~\cite{Shechtman}. 

The second revolution regarded the observation and subsequent theoretical characterization of the quantum Hall 
effect~\cite{QHE,TKNN}. This prompted the realization that phase transitions cannot all be characterized solely in terms of the continuous symmetries they 
break. 
It lead to a new, topological, classification scheme of all non-interacting fermion theories according to discrete symmetries, in what has come to be known as 
the `Tenfold Way'~\cite{AltlandZirnbauer,Tenfold}. 

Recently these two cornerstones of modern condensed matter physics have been shown to be mathematically linked through the ubiquitous Harper 
equation~\cite{Harper}. This equation has long been known to govern electrons on a 2D lattice in the presence of a magnetic field, \emph{i.e.} the 
quantum Hall effect~\cite{Hofstadter}. It was also recently shown to provide a description of 1D quasicrystals when used to describe 
electrons in a 1D lattice with an incommensurate periodically modulated on-site potential~\cite{Kraus2}. Based on the equivalence of the underlying 
mathematical structures of these two systems, quantized transport properties analogous to the quantum Hall conductance were predicted for the quasicrystal, and 
subsequently observed in an experimental realization using optical waveguides~\cite{Kraus1}.

The possibility of ascribing 2D quantum numbers to a system with 1D quasicrystalline order contradicts the prediction of the Tenfold Way, and consequently has 
been widely heralded as evidence of the effectively higher-dimensional origin of quasicrystalline states~\cite{Kraus1,Kraus3,Quandt,Lang,Xu,Ganeshan}. 
This interpretation, however, was recently called into question by Madsen \emph{et al.}, who instead argue that the waveguide experiments involve a second 
degree of freedom which generates a family of quasicrystals~\cite{Brouwer}. The resulting parameter space is therefore 2D, and the topological quantum number 
should be assigned to the entire 2D family of related quasicrystalline states~\cite{Brouwer}.

In this letter we propose that a natural resolution to this disagreement can be found by applying the Harper equation to the description of charge order in 
(quasi-)1D materials. As in the case of optical waveguides, we find that a mean-field description of the charge-ordered state at different 
electronic filling fractions can be labeled by the same set of topological quantum numbers as the 2D quantum Hall effect. We further demonstrate that the 
incommensurate charge-ordered state indeed has a quasicrystalline character. 

In contrast to the optical waveguide implementation, however, the charge-ordered system allows for the interpretation of quasicrystalline order as the limiting 
case of a sequence of different crystalline orders.  
This fact enables us to explicitly identify a second degree of freedom, the phase of the order parameter, which generates a family of 1D charge-ordered states. 
It is this 2D family of states that can be classified by 2D topological quantum numbers. 
In addition, we show that the topological classification of charge-ordered materials is robust against effects beyond the mean-field level, and we propose 
experimental tests of both the topological character of charge-ordered 1D materials and their equivalence to quasicrystals. These tests provide experimental 
access to a novel type of quantized adiabatic particle transport.

\begin{figure}
\centerline{{\includegraphics[width=0.95 \columnwidth]{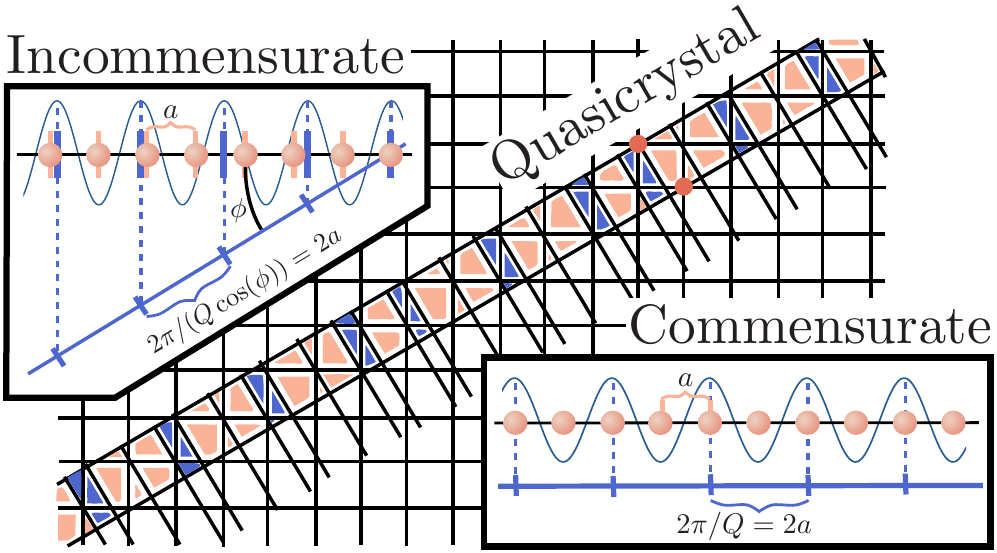}}}
\caption{ {\bf Main:} generating a 1D quasicrystal by projecting a 2D lattice. A pair of parallel lines are overlaid on the square 
lattice, in such a way that they hit only a single pair of lattice points on opposing corners of one of the square unit cells. The lattice points between the 
lines are then projected, and form the sites of the 1D quasicrystal. The inequivalent cells in the quasicrystalline lattice are colored red and 
blue for clarity. {\bf Top Inset:} generating an incommensurate charge-ordered state with period $2\pi/Q$ on an atomic lattice with spacing $a$, by a similar 
projection method. In this case the sequence of atomic sites and maxima of the charge density modulation (red balls and blue tick marks) is quasiperiodic.
{\bf Bottom Inset:} the same method also trivially generates commensurate charge order, when the angle between the two lines is zero.}
\label{fig:projection}
\end{figure}

\emph{Quasicrystals}---Quasicrystals are defined to be quasiperiodic tilings of two (or more) inequivalent unit 
cells~\cite{SnS2}. A quasiperiodic tiling in 1D can be generated from the projection of a regular 2D crystal~\cite{SnS2}.
As shown in Figure~\ref{fig:projection}, a straight line can be drawn in a square 2D lattice, in such a way that it hits exactly one 
lattice point. A second line is then drawn parallel to the first, going through the opposite corner of the unit cell. Whenever a point of the 2D 
lattice falls between the lines, its projection onto the first line forms a site of the 1D quasicrystalline lattice. The projected lattice 
consists of precisely two unit cells, of different sizes. These appear to be randomly tiled, in a never-repeating pattern, but are in fact perfectly 
predictable: the type of unit cell at a given location is known with certainty owing to the periodicity of the 2D parent lattice. This is the 
definition of quasiperiodicity adopted here.

A remnant of the 2D projection underlying the 1D quasicrystal can be seen in its diffraction pattern, which is generated by two 
different reciprocal lattice vectors~\cite{SnS2}. This argument was recently employed to explain the observation of an integer number of edge 
states in 1D quasicrystals realized in optical waveguide arrays~\cite{Kraus1}. The number of edge states is a direct measure of the Chern number 
$C_{1}$ of a system, which depends only on the system's topological character~\cite{TKNN}. 
According to the accepted classification of fermion systems with no assumed symmetries,
Chern numbers that can take any integer value, as in the waveguide experiments, exist in systems with even 
spatial dimensions~\cite{AltlandZirnbauer}. Odd-dimensional systems have only trivial topological phases, without edge modes~\cite{HasanKane,Tenfold}.

An alternative explanation of the seemingly paradoxical presence of 2D Chern numbers in 1D quasicrystals builds on the assertion 
that the waveguide experiments involve two degrees of freedom rather than one~\cite{Brouwer}. 
The additional parameter relates distinct but locally isomorphic quasicrystals, where two quasicrystals are defined to be locally isomorphic if and only if 
every finite sequence of unit cells which appears in one also appears in the other~\cite{SnS2}. The experimentally observed edge states and their related 
quantized transport properties are then argued to relate to the entire 2D family of quasicrystals, rather than to any one of its 1D 
members~\cite{Brouwer}.

\emph{Charge Order}---The argument about the effective dimensionality and topological character of quasicrystals can be straightforwardly reformulated in terms 
of properties of charge-ordered (quasi-)1D materials. In such systems chains of atoms spontaneously develop a modulation in their electronic 
density that does not coincide with the periodicity of the underlying ionic lattice. They include well-studied materials like NbSe$_3$, KCP, and TTF-TCNQ, 
which all contain weakly coupled 1D chain structures~\cite{Wilson,Hodeau,Eagen,Denoyer,Jerome}. 

The connection with quasicrystals can be made using a simple projection method. Consider a 1D atomic lattice with spacing $a$ and a sinusoidal modulation of 
charge density with period $2 \pi / Q \neq a$, as indicated in the inset of Figure~\ref{fig:projection}. For incommensurate charge order, the sequence of 
atomic 
sites and maxima in the charge modulation along the chain form a seemingly random, never-repeating, pattern.
It can be easily seen however that the pattern is perfectly predictable as it arises from the projection of a second 1D lattice with lattice spacing $n a$, 
where $n$ is the lowest integer number larger than $2 \pi / Q a$. This second lattice can always be rotated to a point where the projections of its lattice 
sites onto the original line coincide with the maxima of the charge modulations (see Figure~\ref{fig:projection}). The sequence of sites and charge maxima in 
the original lattice is therefore quasiperiodic. 
The combined system can be interpreted as a quasicrystal, which could in principle be observed in scanning tunneling microscopy experiments.
By continuously shifting the charge order with respect to the atomic lattice, a continuous set of locally isomorphic quasicrystals is generated. 

The emergence of charge-ordered states can be most easily understood in a model of spinless fermions hopping on a 1D lattice, in the presence 
of nearest-neighbor density-density interactions (strength $h$)~\cite{Gruner}. Such a 1D system is always unstable towards a spontaneous 
modulation of its local electronic density. This can be described at the mean-field level by introducing the expectation value 
$\Delta_{Q}\left(h\right)=2h \sum_k \left\langle \hat{c}_{k+Q}^{\dagger}\hat{c}_{k}^{\phantom{\dagger}}\right\rangle $, where $Q$ is a wave vector connecting 
the two points on the Fermi surface of the 1D band structure. Using this decoupling, the mean-field Hamiltonian can be written as:
\begin{align}
\hat{H}=\sum_{0 \le k < 2 \pi / a} \left\{ \frac{1}{2}\epsilon_{k}^{\phantom \dagger}\hat{c}_{k}^{\dagger}\hat{c}_{k}^{\phantom \dagger}+\Delta_{Q}^{\phantom 
\dagger}\hat{c}_{k}^{\dagger}\hat{c}_{k+Q}^{\phantom \dagger}+H.c.\right\}
\label{eq:H_MFT}
\end{align}
where $\epsilon_{k}=2t\left(1+\cos\left(k a\right)\right)-\mu$ describes the bare band structure resulting from the hopping integral $t$ and chemical potential 
$\mu$, and the 
charge density order parameter can be written as $\Delta_{Q}=\left|\Delta_{Q}\right|\exp\left(i\theta\right)$. If the chemical potential is tuned to a rational 
fraction $p/q$ of the bare bandwidth $4 t$, the charge order in the ground state of the full mean-field Hamiltonian in Equation~\eqref{eq:H_MFT} will have 
period $q a$. Working in a reduced Brillouin zone of length $2\pi/qa$, the Hamiltonian can then be rewritten as:
\begin{align}
\hat{H}&=\sum_{0 \le k < 2\pi / qa} \left(\hat{c}_{k+Q}^{\dagger},\hat{c}_{k+2Q}^{\dagger}  \ldots \hat{c}_{k+qQ}^{\dagger}\right)H_{k}\left(\begin{array}{c} 
\hat{c}_{k+Q}\\ \hat{c}_{k+2Q}\\ \vdots\\ \hat{c}_{k+qQ} \end{array}\right) \notag \\
~&~\notag\\
H_{k} &= \left(\begin{array}{cccccc}
\epsilon_{k+Q}^{\phantom *} & \Delta_{Q}^{\phantom *} & 0 & \ldots & 0 & \Delta_{Q}^{*}\\
 \Delta_{Q}^{*} & \epsilon_{k+2Q}^{\phantom *} & \Delta_{Q}^{\phantom *} & 0 & \ldots & 0\\
0 & \Delta_{Q}^{*} & \ddots & \ddots &  & \vdots\\
\vdots & 0 & \ddots &  &  & 0\\
0 & \vdots &  &  &  & \Delta_{Q}^{\phantom *}\\
 \Delta_{Q}^{\phantom *} & 0 & \ldots & 0 & \Delta_{Q}^{*} & \epsilon_{k+qQ}^{\phantom *} \end{array}\right).
\label{eq:H2}
\end{align}

\begin{figure}
\centerline{{\includegraphics[width=0.95 \columnwidth]{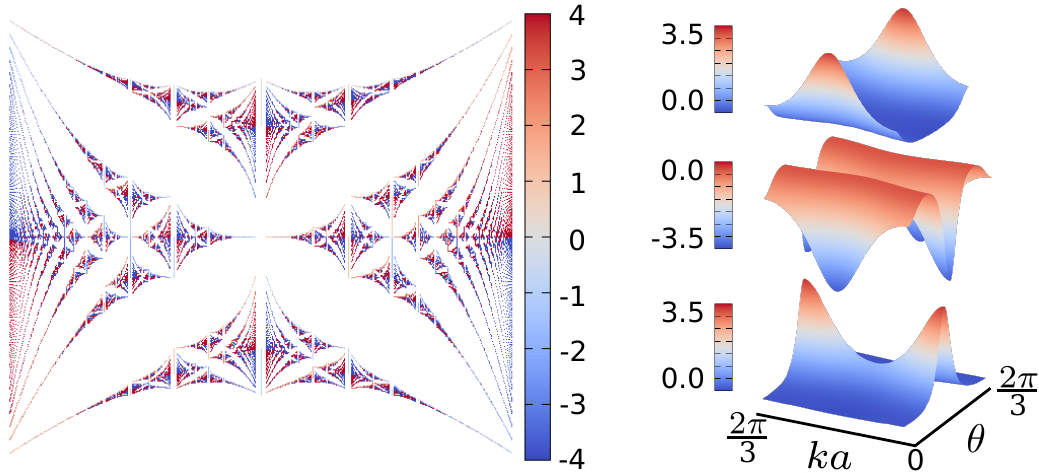}}}
\caption{{\bf Left:} allowed energies (range $\left[-2t,2t\right]$) versus filling fraction $p/q\in\left(0,1\right)$, $q\in\left[2,50\right]$ in the mean-field 
Equation~\eqref{eq:H2}, with $|\Delta_Q|/t=1$. The image is closely related to the spectrum of Hofstadter, describing allowed electron energies in the integer 
quantum Hall effect~\cite{Hofstadter}. The color of each sub-band indicates the sum of the Chern numbers $C_{1}$ up to and including that sub-band (the scale 
has been truncated at $\pm4$ for clarity). {\bf Right:} the Berry curvatures (color coded, images displaced vertically for clarity)
for the sub-bands at $1/3$ filling. The integrals over these surfaces give (top to bottom) $1,-2,1$ in units of $2\pi$, which are the corresponding $C_{1}$ 
values for each sub-band.}
\label{fig:Butterfly}
\end{figure}

For the case $\left|\Delta_Q\right|/t=1$ this is precisely the matrix form of Harper's equation, applied by Hofstadter to the 2D electron gas in the 
presence of a magnetic field~\cite{Hofstadter}. A plot of the eigenvalues against filling fraction has come to be known as Hofstadter's butterfly 
(Figure~\ref{fig:Butterfly})~\footnote{Our image is a slight variation on Hofstadter's since we discard non-coprime fractions. An example is $p/q=1/2$ which 
has $2$ bands in our case but is a solid line in Hofstadter's. Our color coding also implies a slight variation on statements made for example in 
Reference~\cite{Bloch}: for $1/4$ filling the central bands touch, so their Chern numbers should be added. We note that in our case higher harmonics of the 
charge order would couple the relevant sub-bands and cause a splitting.}. The eigenvalues of the matrix in Equation~\ref{eq:H2} constitute the possible 
energies of the system after the charge order has been established. In the presence of a charge modulation with period $q a$, the energies form $q$ sub-bands, 
separated by energy gaps~\cite{Gruner}.

The contribution of each filled sub-band to the overall conductivity of the system is quantized in units of $2\pi e^{2}/\hbar$~\cite{TKNN}. The quantization of 
conductance can be understood by realizing that the phase, $\theta$, signifying the position of the charge modulation relative to the atomic lattice, is not 
fixed by the mean-field solution of Equation~\eqref{eq:H2}. The band structure can thus be drawn within a 2D space spanned by $\theta$ and $k a$, 
as indicated on the right of Figure~\ref{fig:Butterfly} for the case $p/q=1/3$. Each of these 2D sub-bands can be assigned a topological quantum 
number, the Chern number $C_{1}$, by integrating the Berry curvature over this 2D space. In this specific problem $C_{1}$ can also be 
found by solving a Diophantine equation~\cite{ChangNiu1,TKNN}. The quantized conductance is then the sum of the Chern numbers for all occupied 
sub-bands~\cite{TKNN}. In the quantum Hall effect, the Hall conductivity $\sigma_{xy}$ is given by $(2\pi e^{2} / \hbar) C_{1}$. In the present case, the 
response function associated with the topological invariant is instead a `quantized adiabatic particle transport'~\cite{QAPT,Thouless}: adiabatically cycling 
$\theta$ through a full period 
results in the transfer of an integer number $C_{1}$ of electrons across the length of the 1D chain. Since $\theta$ describes 
the position of the charge modulation with respect to the atomic lattice, this cycling of $\theta$ corresponds to rigidly sliding the charge-ordered state 
through a single wavelength.

The Chern numbers for the commensurate charge-ordered patterns which make up the left of Figure~\ref{fig:Butterfly} are indicated by its color scale. Only 
commensurate charge density modulations appear in the figure, but properties of the incommensurate charge-ordered states can be inferred from its large-scale 
structure. In particular, approaching an irrational value of the filling fraction as the limit of a series of rational fillings, it is clear that there must be 
large gaps 
(for example at the wings of the butterfly) even for incommensurate charge density modulations. We can then use the self-similarity of the fractal to see that 
any incommensurate state at irrational filling fraction $\eta$ in Figure~\ref{fig:Butterfly} must have the same nonzero sum of 
Chern numbers for its filled bands as the state with commensurate order at the rational filling $p/q=\eta-\epsilon$, with $\epsilon\rightarrow0$, which 
approaches $\eta$ 
arbitrarily closely. For example, a close 
inspection of the underside of the butterfly's lower left wing reveals a constant lining of $C_{1}=1$ for all commensurate states filled up to the wing, which 
must therefore also apply to the incommensurate states interpolating between them.

The Chern numbers characterizing the quasicrystalline state can take on any integer value depending on the value of the (irrational) filling fraction. This 
observation does not contradict 
the usual topological 
classification, as the Berry curvature giving rise to the Chern numbers is integrated over both $k$ and $\theta$, so that the relevant parameter space is 2D. 
The quantized transport and the corresponding edge states are properties of the entire 2D family of locally isomorphic charge-ordered states related by 
translations over $\theta$, in direct analogy to the case of the optical waveguide experiments~\cite{Kraus1,Brouwer}.

\begin{figure}
\centerline{{\includegraphics[width=0.95 \columnwidth]{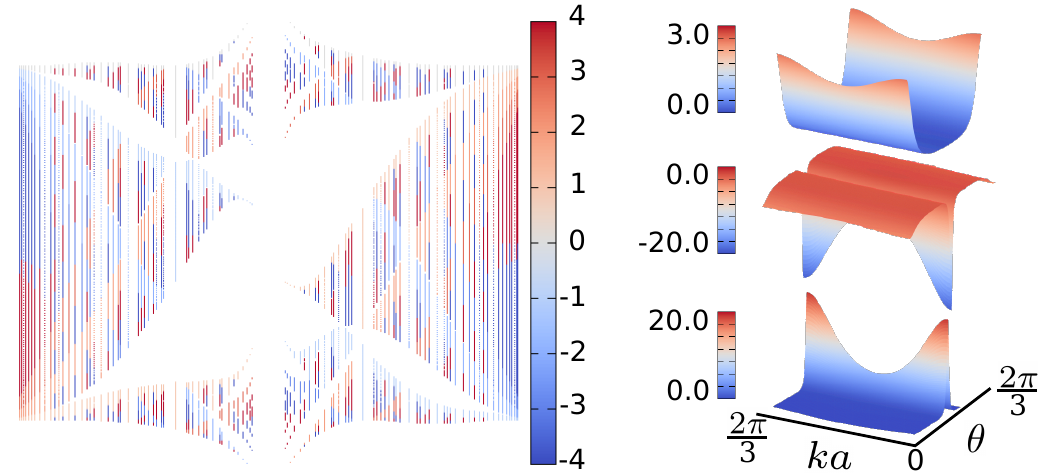}}}
\caption{{\bf Left:} the allowed energies (scale truncated at $\left[-2.5t,2.5t\right]$) as a function of filling fraction $p/q\in\left(0,1\right)$, 
$q\in\left[2,18\right]$ 
for the self-consistent solution of Equation~\eqref{eq:H_MFT} with $h=t$. Each sub-band is adiabatically connected to the corresponding sub-band in 
Figure~\ref{fig:Butterfly}. This implies that the sums of Chern numbers of filled sub-bands, indicated by the color scale, are the same. {\bf Right:} the 
Berry curvatures (color coded and displaced for clarity) for the three sub-bands at $1/3$ filling. Although the shapes are different to those in 
Figure~\ref{fig:Butterfly}, the integrals over them remain $1,-2,1$ in units of $2\pi$.}
\label{fig:sc-Butterfly}
\end{figure}

\emph{Experimental Realization}---In the analysis of the spectrum of Equation~\eqref{eq:H2} we imposed a fixed value of 
$\Delta_{Q}\left(h\right)$, and did not take into account the self-consistency condition $\Delta_{Q}\left(h\right)=2h \sum_k \left\langle 
\hat{c}_{k+Q}^{\dagger}\hat{c}_{k}^{\phantom{\dagger}}\right\rangle $. 
Although the self-consistency requirement can affect the sizes of the different gaps in Figure~\ref{fig:Butterfly}, the topology of the sub-bands should not be 
affected, as none of the gaps can be closed up entirely. 
This implies that the Chern numbers characterizing the contribution of each sub-band to the adiabatic transport will be protected. 
Solving Equation~\eqref{eq:H_MFT} self-consistently, we obtain the allowed energies and Chern numbers displayed in 
Figure~\ref{fig:sc-Butterfly}~\cite{ChangNiu1,TKNN}. As expected, 
the value of $C_{1}$ associated with each sub-band is unaltered by deformations of $\left|\Delta_{Q} \left(h\right)\right|$. 
The self-consistent solution at a given interaction strength $h$ therefore has the same topology as the standard Hofstadter case of 
Figure~\ref{fig:Butterfly}.

Real materials cannot maintain the infinite fine structure of a fractal band. Instead, the gaps separating consecutive sub-bands will be closed 
when they become of the order of the effective energy scale set by local impurities, disorder, or lattice defects. When this 
happens, the Chern numbers of the two merging bands are added together to yield the Chern number describing the combined sub-band~\cite{Bloch}. The larger gaps 
that correspond for example to the wings of the butterfly may be expected to survive in clean enough systems at low temperatures. For 
the corresponding filling fractions, the sum of Chern numbers must then still be nonzero, even for incommensurate charge-ordered states in the presence of
defects and disorder.

A second effect of the presence of impurities in real materials is their ability to pin the charge order in place, inhibiting the freedom of the modulated 
pattern to slide along the atomic lattice. It is well established, however, that an applied voltage may be 
used to overcome the pinning potential in incommensurate charge-ordered materials, allowing them to conduct even in the presence of pinning 
centers~\cite{Bardeen,Fukuyama76,LeeRice,Gruner}. 
The current produced in such a way includes both the effects of adiabatic particle transport originating in the band topology, and the effects of non-adiabatic 
band-mixing. Since we are here only interested in the former effect, we propose an alternative way of accessing the quantized particle transport in both 
commensurate and incommensurate charge-ordered states, based on the use of atomic condensates in optical lattices \footnote{We 
address the issue of quantized adiabatic particle transport in quasi-1D materials in an upcoming paper.}. 
Such systems are more readily controllable and defect-free than 1D chains in actual materials. Indeed, the Harper equation has recently been realized in a cold 
atom 
setup~\cite{Bloch}. Impurities (including the confinement potential) will inevitably exist, but we propose to use them to our advantage. 
An impurity can be simulated by altering the potential on a single site, providing for example an attractive center which locks one of the maxima of the charge 
density modulation 
into place. Provided that this intentional pinning does not disrupt the large-scale gap structure, the topology of the sub-band remains unaffected. By 
manipulating the impurity location, the charge density modulation can be dragged along the optical lattice, continuously cycling the value of 
$\theta$ along its entire range. As the corresponding charge-ordered pattern moves through one wavelength, it should be possible to measure the transport of an 
integer number ($C_{1}$) of electrons across the length of the optical lattice.

\emph{Conclusions} - In this letter we have shown the equivalence between families of 1D locally isomorphic quasicrystals and a mean-field model of 
incommensurate charge order in 1D materials. Both exhibit the same set of topological quantum numbers normally ascribed to 
2D systems, in agreement with recent results on optical waveguide experiments~\cite{Kraus1,Kraus2}. Using the self similarity of the fractal pattern of allowed 
energies to interpret  incommensurate charge-ordered materials as the limit of a series of commensurate cases, it becomes clear that these Chern numbers arise 
from an integral of 
the Berry curvature over a 2D space covering the entire family of locally isomorphic charge-ordered states. 
It is therefore the family of 
modulation patterns, rather than any individual quasicrystalline state, which is characterized by a 2D topological quantum number~\cite{Brouwer}.

We find no conceptual difference between the topologies of the sub-bands for commensurate and incommensurate charge order, in the sense that the properties of 
an incommensurate charge-ordered state can be determined from the limit of a series of commensurate phases using the large-scale structure of the 
Hofstadter spectrum. The topological properties of the system survive in the full self-consistent solution of the model, and are robust against the inclusion 
of 
weak fluctuations, defects and impurities. The large-scale structure of the Hofstadter spectrum guarantees the 
possibility of quantized adiabatic particle transport in charge-ordered materials, as long as the gap size remains nonzero. 

Finally, we propose an experimental test of the topological properties of charge-ordered materials by exploiting the possibility of directly controlling the 
phase of a charge density modulation in atomic condensates in optical lattices. We predict that dragging the phase through a full period by this method 
will lead to the transfer of a quantized number of electrons across the system, providing a novel standard of conductance.


%

\end{document}